\documentclass[preprint]{vgtc}            

\preprinttext{\parbox{\textwidth}{This work has been submitted to the IEEE for possible publication. Copyright may be transferred without notice, after which this version may no longer be accessible.}}

\vgtccategory{Research}

\title{Seeing through the Eyes of AI: Situated Explainability in Augmented Reality}

\author{Ana Stanescu\thanks{e-mail:ana.stanescu@adelaide.edu.au}
\\ \scriptsize Adelaide University %
\\ \scriptsize Graz University of Technology %
\and Lucchas Ribeiro~Skreinig
\\ \scriptsize Graz University of Technology %
\and Tobias Langlotz
\\ \scriptsize Aarhus University %
\and Stefanie Zollmann
\\ \scriptsize Aarhus University %
\and Peter Mohr
\\ \scriptsize Graz University of Technology %
\and Dieter Schmalstieg
\\ \scriptsize University of Stuttgart %
\and Mark Billinghurst
\\ \scriptsize Adelaide University %
\and Denis Kalkofen\thanks{e-mail: kalkofen@tugraz.at}
\\ \scriptsize Graz University of Technology %
}

\abstract{%
Explainable Artificial Intelligence (AI) enables humans to understand and interpret decisions of AI models. Instead of having a black box, explainability supports humans in understanding AI models' behavior. Existing explainable AI approaches often present explanations on 2D displays using pre-recorded data, requiring users to relate the displayed information back to the physical objects and real world locations involved in a model’s decision. Users are forced to decouple data exploration and capture from AI model interpretation. For AI systems that work within physical environments, this separation can make explanations difficult to interpret in context. We propose using Augmented Reality (AR) to enhance the understanding of AI models by enabling spatial explainability information directly in a user’s workspace, in real time, as they explore the world. We show how known explainability methods can be applied in AR and provide insights into user experiences with such an application. 
}


\teaser{
  \centering
  \includegraphics[width=\linewidth, alt={A view of a city with buildings peeking out of the clouds.}]{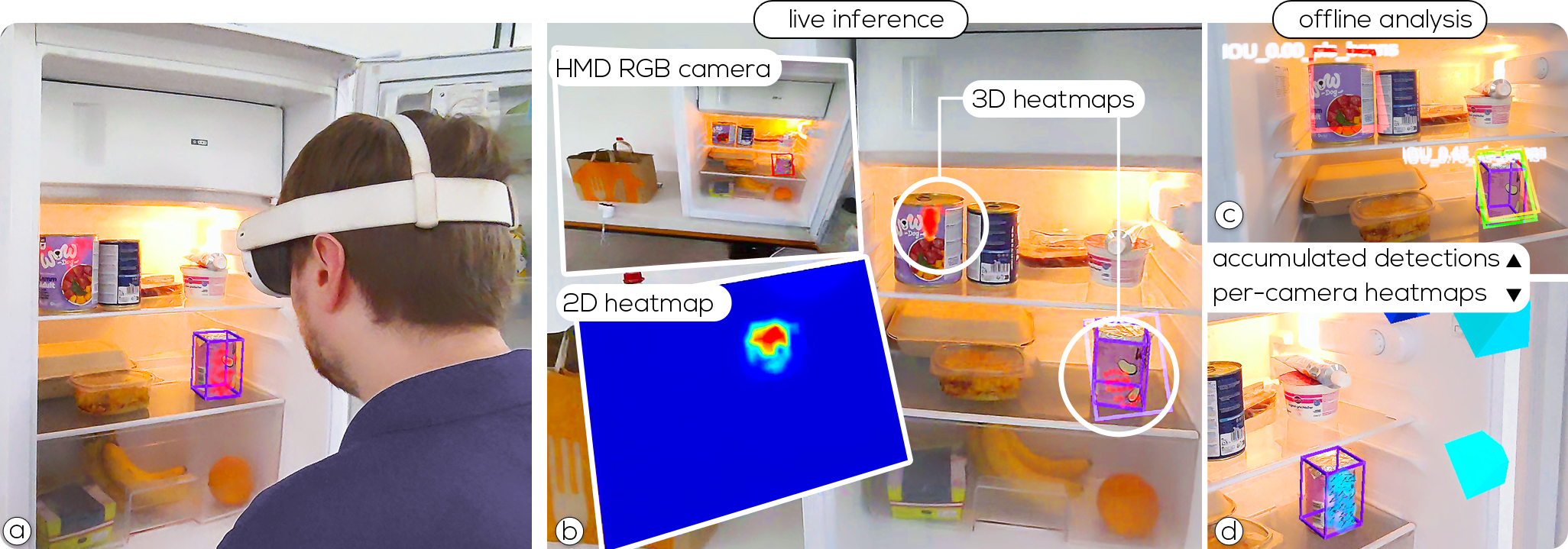}
  \caption{
  	(a) A user running an AR application using object detection in their fridge. (b) The situated explainability framework displays information about the network's decisions directly into the user's surroundings at runtime, with information such as weighted activations for the detected class (a can of beans on the bottom shelf) also appearing on the pet food can on the top shelf, because of visual similarity causing confusion in the detector. (c) The user can then see the aggregated detections over time and (d) the corresponding camera poses for the viewpoints where the detections occurred can be inspected later. 
  }
  \label{fig:teaser}
}

\graphicspath{{figs/}{figures/}{pictures/}{images/}{./}} 
\usepackage{tabu}                      
\usepackage{booktabs}                  
\usepackage{lipsum}                    
\usepackage{mwe}                       
\usepackage{mathptmx}                  

\begin{document}

\firstsection{Introduction}

\maketitle

This paper introduces the concept and first exploration of \emph{situated explainability}, an approach that uses Augmented Reality~(AR) to deliver explainable Artificial Intelligence~(AI) in the space around the user. 
Advances in AI and Machine Learning (ML) models have enabled mobile personal devices such as mobile phones and smart glasses to create a better understanding of the user’s environment and spatial context. These visual perception models often use computer vision (CV) techniques and neural networks processing data captured from the user's perspective (egocentric data). 

Applications of this type of \textit{physical AI} range from simple object detection, to AR instructions and maintenance systems, where detected objects are used to provide contextual guidance to the user \cite{stanescu2023state}. Physical AI is also a key component of AR browsers augmenting the environment with digital information \cite{langlotz2014next}, in medical applications \cite{song2025enhancing}, or gaming experiences \cite{wang2020}. More generally, physical AI is essential for any application that benefits from a reliable, detailed understanding of the spatial surroundings from visual data. 

However, AI methods are prone to errors and dataset bias \cite{kim2019learning}, and their decision-making is typically accompanied by little or no explanation ('black-box models'). This lack of transparency can be problematic in applications where incorrect detections may lead to user errors and can reduce confidence in the system. AI explainability aims to address this by studying methods that make ML models more interpretable, allowing users to better understand what is learned and how decisions are made. Previous work has shown that explanations can improve user understanding \cite{alqaraawi2020evaluating, wang2021explanations, rong2023towards}, and it is generally agreed that explainability is a crucial aspect for the responsible use of AI and can affect trust in the model's decisions \cite{rong2023towards}. 

Specifically, when learning visual features, methods based on \textit{class activation mapping} (CAM)~\cite{zhou2016learning}, such as Grad-CAM~\cite{selvaraju2017grad}, are commonly used for post-hoc explainability. CAM highlights class-relevant regions to provide meaningful insight into whether a model is focusing on relevant object features, as heatmaps. CAM methods can help identify dataset biases, object localization behavior, discriminative feature usage, spurious correlations, systematic misclassifications, robustness, and class-specific reasoning differences, making them popular in CV~\cite{jiang2021layercam,  zhou2016learning}, medicine~\cite{sun2025explainable} or industry \cite{tzionis2025review}. CAM methods have also been used for other data types than images, such as time series~\cite{van2024harnessing, theissler2022explainable}, spectrogram data~\cite{carofilis2023improvement}, or 3D data from human skeletons~\cite{li2024gt}. 

However, CAM-based methods are typically presented as overlays on previously captured images or within a separate two-dimensional camera view. Even when applied to a live video stream, users must relate locations in the displayed image back to objects and regions in the physical environment. AR can reduce this separation by spatially registering explanations with their physical referents, allowing users to inspect model behavior directly in the context in which the model operates.

We propose the concept of \textit{Situated Explainability} that uses AR as an interface to visually inspect physical AI relying on visual imagery. Rather than inspecting a static sample from a test set on a desktop computer, situated explainability provides critical insights into the decision-making of visual AI models directly in a user's physical environment\cite{white2009sitelens}, enabling immediate, live feedback and a deeper understanding of visual AI models in their actual spatial context. As such, situated explainability can be defined as \textit{the visualization of AI explainability methods directly in the physical world, with the explanations being spatially registered to the AI model's inputs, and provided at runtime, synchronized with decisions of the model}.

In this paper, we introduce the first approach that combines visual explainability of AI with an AR interface that provides important live feedback and goes beyond textual explanations and use of pre-recorded data. Our findings contribute to the area of explainability of AI models but also make a contribution to the growing areas of user interfaces for explainable AI. Specifically, our contributions are (1) the \textbf{introduction of the situated explainability concept}, (2) its \textbf{ first implementation}, that demonstrates immersive exploration of visual models and their direct interaction with visual features in place, (3) presenting the \textbf{results of a user study with participants with AI background} using situated explainability as implemented in our prototype. Our results highlight a more engaging experience at the cost of a higher cognitive workload (as compared to traditional explainability workflows) and show that participants clearly preferred \emph{situated explainability} over traditional tools.

\section{Related Work}

The use of AI and especially physical AI models is increasingly relevant for mobile and wearable computing applications, specifically those using AR as an interface. For example, visual detection has been employed in AR instructions to detect objects, object poses or states, enabling automatic AR tutorial applications~\cite{stanescu2023state, stanescu2024error, schieber2024asdf, zhou2020fine, su2019deep}. A Large Language Model\,~(LLM) can detect states and provide feedback \cite{srinidhi2024xair} in pervasive AR interfaces \cite{grubert2016towards}. XR-Objects \cite{dogan2024augmented} uses a detection-based pipeline to identify objects and display contextual information in the user's view. Search tasks \cite{raikwar2024beyond} use detectors to identify relevant objects around a user and use AR to help find them. 

Recently, foundation models that can receive image data as input have gained popularity, such as LLaVa~\cite{liu2023llava}, GPT-4V~\cite{gpt4v} or Gemini~\cite{team2023gemini}. For real-time AR applications, the choice between standard object detectors and foundation models involves a trade-off between efficiency and flexibility. In their current form, these models are generally larger, slower, and more complex to deploy, suffering from higher latency, increased power consumption and less stable frame-to-frame predictions. Standard object detectors, e.g., YOLO~\cite{wang2024yolov10} or RT-DETR~\cite{zhao2024detrs}, are optimized for fast inference, offering low latency and efficient device deployment, as required for AR systems.

Explainability helps to understand the inner workings of such ``black-box'' AI models. However, it is usually studied offline on individual samples. Existing work in AR has explored presenting explanations through text, based on AI output \cite{xu2023xair}. In contrast, we embed explainability visualizations directly into the user's physical environment, generating fine-grained saliency maps in real time and registering them spatially to the objects and regions seen by the AI model. These visualizations allow users to inspect what the AI is attending to \emph{directly in the environment where it operates, rather than through a separate 2D or virtual interface}.

\subsection{Explainability visualizations}

In ante-hoc explainability methods, models are designed to be inherently interpretable, while post-hoc methods focus on explaining the predictions of already trained black-box models \cite{bhati2025survey}. Well-known post-hoc methods include SHAP~\cite{lundberg2017unified} to estimate feature importance, and activation-based approaches such as GradCAM~\cite{selvaraju2017grad} with its variants, including EigenCAM~\cite{muhammad2020eigen} and LayerCAM~\cite{jiang2021layercam}, which provide more detailed visualizations of network activations. These heatmap-based visual approaches~\cite{la2023state} have been applied in domains where it is essential to understand network decisions, such as medicine~\cite{tinauer2022interpretable}. Initially introduced for convolutional neural networks, CAM methods have been extended to a wide range of neural architectures, including transformers~\cite{li2024gt}.

To provide simple explanations for everyday users, natural language has been used in addition to CAM methods. For example, a GradCAM visualization in an image helps to identify AI-generated art \cite{castellano2024using}. However, LLM-generated explanations can contain hallucinations or inaccurate information, which affects the reliability of the explanations.

In practice, post-hoc explainability methods are mostly used in offline analysis. Interactive visual analytics systems such as explAIner \cite{spinner2019explainer}, ConceptExplainer \cite{huang2022conceptexplainer}, ScrutinAI \cite{haedecke2023scrutinai} and GALORE \cite{wang2023generalized} provide interfaces for exploring explanations of trained neural networks. However, these systems are designed for conventional desktop environments, where users inspect explanations through 2D visualizations on a monitor using mouse-and-keyboard interaction. They are primarily intended for post-training analysis of pre-recorded data rather than real-time or spatial workflows.

\begin{figure*}[t]
   \centering
   \includegraphics[width=0.99\linewidth]{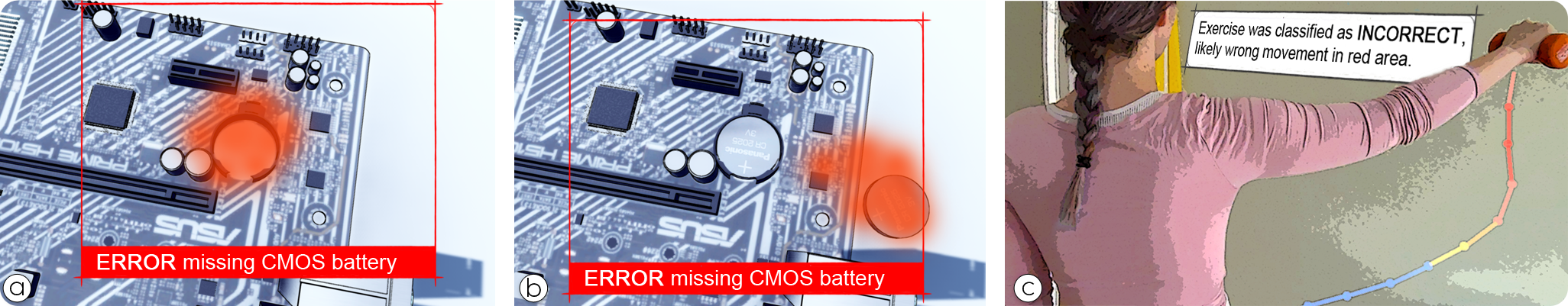}
  \caption{Envisioned applications for live situated explainability (proof-of-concept images): In an industrial inspection task visual features of a chip lead to an AI model suggesting a missing battery error. (a) The area with the missing battery is highlighted by CAM, meaning the correct features were learned. (b) The network learns that a battery next to the board means that it is not mounted, which is not necessarily correct. (c) In a physiotherapy tutorial, situated explainability can highlight parts of a 3D motion trajectory that influenced the prediction of the class ``incorrect''. 
  }
    \label{fig:envisioned_applications}
    \vspace{0em}
\end{figure*}

\subsection{Explainability in AR and VR}

AR is a promising environment for AI explainability, as AR explanations can be seamlessly spatially coupled to predictions related to the very objects, regions, and actions they explain, making it a natural application domain for explainable AI \cite{mendez2023does}. However, research on explainability in AR is still in its early stages.

Xu et al. \cite{xu2023xair} created a design framework for explainable AI in AR, and conducted a broad user study to identify requirements and present conceptual applications of explainability in AR. They focused on textual explanations and assumed existing XAI techniques rather than integrating explainability algorithms into the AR pipeline. Similarly, PILAR \cite{kundu2025pilar} is an AR framework that employs an LLM to generate context-aware textual explanations of AI recommendations.

Further applications have been proposed that allow a user to interact with an LLM to ask for explanations about cybersickness in VR \cite{kundu2025relaxvr} or about analyzing user behaviors in XR \cite{kim2025explainable}. While LLM-based explanations can provide personalized and human-understandable feedback, they rely on a separate language model to generate natural language from model output. Therefore, the faithfulness of these explanations to the underlying AI model is not guaranteed, and the work does not explicitly evaluate or mitigate potential hallucinations.

De Heuvel et al. \cite{de2025immersive} visualize attribution scores for robot navigation based on reinforcement learning by highlighting relevant virtual objects in VR. Their user study shows that users can better understand and predict robot behavior.  
While also providing real-time visual explanations in an immersive environment, they represent attribution at the level of entire semantic objects for a robot task in a fully virtual environment, while our work adds fine-grained visual explanations to the user's real, physical environment using AR.

\subsection{AR debugging and data collection}

AR has also been proposed as a debugging tool. ARGUS is a debugging application for AR sessions \cite{castelo2023argus} aimed at developers, to visualize statistics about AI model performance. The tool is designed to focus on data collection without offering further explainability or interpretation, this being left to the developer. ARCHIE \cite{lehman2020archie} is a framework for testing AR applications in the wild, combining user feedback with system-state logs to help developers reproduce and diagnose runtime issues.

ARCADE \cite{ganj2026ar} is a framework for evaluating CV models in AR scenarios. The system replays recorded AR captures, allowing developers to compare model outputs in tasks such as object placement, occlusion rendering, and point cloud visualization, focusing on depth and lighting estimation models. Unlike our work, ARCADE is intended for model evaluation rather than for real-time explainability. Xu et al. \cite{xu2024multimodal} let AR users provide object-level annotations to incrementally train and personalize AI models. Live annotation for object detection has also been proposed for AR \cite{qian2022arnnotate, taheritajar2025scalable}.

Unlike these systems, our work focuses on understanding the decision-making process of AI models directly in the AR environment through real-time, using spatially registered visual explanations that also enable informed data collection.

\section{Situated Explainability}
\label{sec:situated_explainability}

In this work, we are using AR to bring explainability algorithms originally intended for offline inspection into the user's environment. Explanations are generated in real time and spatially registered to the objects they explain, enabling the user to interactively explore the behavior of the AI model. When models underperform, they can be inspected using situated explainability, and also improved by collecting additional data from the user's surroundings. In the following, we include scenarios that would benefit from our concept (also see Figure \ref{fig:envisioned_applications}). 

\textbf{Checking for faults and reliability}: With the proposed concept, one can check whether the AI is focusing on the relevant discriminative parts. For example, in an industrial quality inspection, the 3D CAM maps could help a technician confirm that defect detection models rely on true fault regions instead of lighting or texture variations. The technician can move around an object and inspect it from multiple viewpoints in AR to check the stability of the detection.

\textbf{Understanding of dataset biases}: In the retail field, visualizations can assist in verifying that models identify products based on their features rather than background or packaging biases. For example, for an AR assistant that helps with shopping or recipes, in cases where the user does not understand why a recommendation occurs, the visualizations could reveal which part of the object is relevant for the network. The user can then directly adjust their view on-the-fly to get other recommendations if needed.

\textbf{AI model testing and comparison}: During model development and debugging, live situated saliency methods can help practitioners understand whether a network has learned relevant visual features or dataset shortcuts. Data engineers can change model architectures and visualize the output live in a controlled test environment, with AR allowing them the flexibility to vary the live conditions.

\textbf{Explaining and debugging classification of motion data}: CAM methods can also be used on other data besides images, and with various network architectures. For example, for motion data, one can visualize which parts of the motion are more discriminative. This can be applied to motion trajectories over time (Figure~\ref{fig:envisioned_applications}c).

\textbf{Spatial understanding}: In an AR re-localization task using an AI model, in-situ explainability can be shown to immediately highlight which features of the scene were used. This visualization would allow the user to move around and visually confirm that important regions for re-localization are considered relevant for the AI model.

\textbf{Quick situated decision-making}: With traditional explainability methods that run on pre-recorded data, users cannot quickly react based on discovered learned flaws. In contrast, with live AR integration, one can quickly understand network behavior and decide how to proceed with the AI model.

\textbf{Checking for model sensitivity}: While inspecting scene parts that are relevant for the AI models, users can occlude or move them, and instantly observe the model behavior in terms of accuracy, recovery, and changes in the explainability visualizations.

\begin{figure}[!b]
   \centering
   \includegraphics[height=2.95cm]{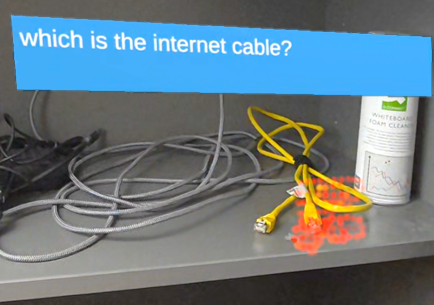}
   \includegraphics[height=2.95cm]{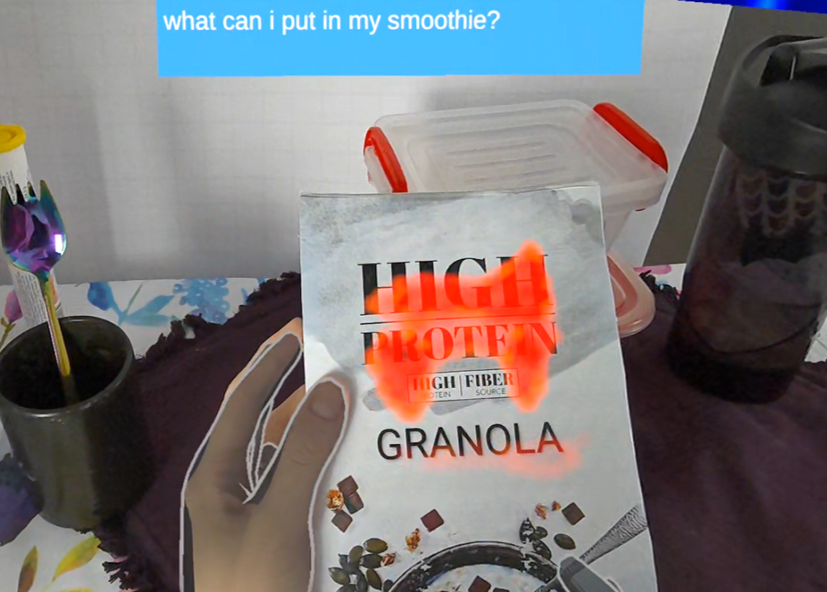}
  \caption{Examples from a situated explainability implementation with a VLM (PaliGemma mix), allowing the user to ask questions in real time, visualizing the VLM attention overlaid on the 3D world.}
    \label{fig:example_vlm}
    \vspace{0em}
\end{figure}

\begin{figure*}[t]
   \centering
   \includegraphics[width=\linewidth]{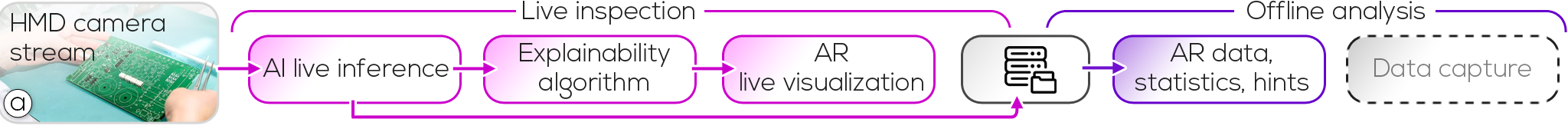}
      \caption{Pipeline of the framework. The user runs live inference in their environment, using the live image feed from their headset. An explainability module is run and the result is sent back to the headset, where it is shown to the user as AR live visualization, or accumulated and later used for offline analysis. Then, the system supports AR offline analysis and optional data recording.}
    \label{fig:pipeline}
    \vspace{0em}
\end{figure*}

\section{Application}

\label{sec:application}

To exemplify the concept of situated explainability in AR, we implement a proof-of-concept AR framework (shown in Figure \ref{fig:pipeline}). As input, we use the live image stream of an AR headset. In \emph{live mode}, a neural network processes images, and an explainability algorithm extracts relevant information, which is displayed to the user in real-time. In \emph{offline mode}, data from prior live mode captures is further processed in an offline step, and analytics are displayed. Using our novel framework and the gained insights, a user can capture more data to re-train the network in the future. Next, we describe each component and its features.

\begin{figure}[b]
   \centering
   \includegraphics[width=0.99\linewidth]{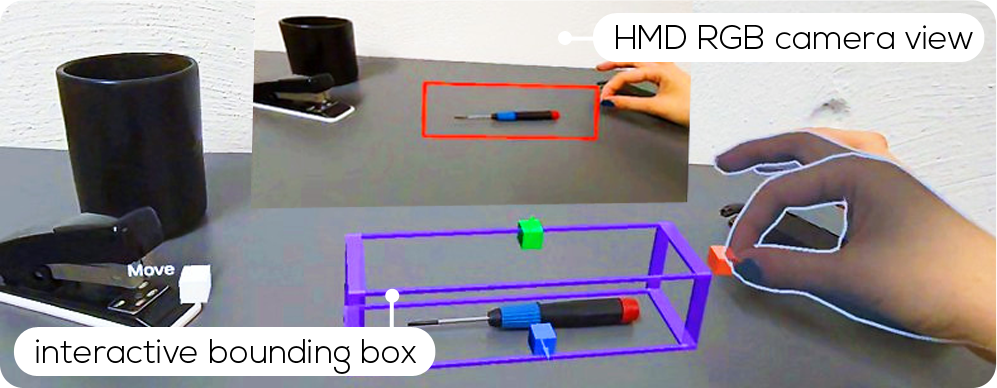}
  \caption{AR object annotation: The user can place 3D labels in the scene for automatic data labeling and checking for the prediction correctness. The 3D labels are projected onto the image plane.}
    \label{fig:labelling}
    \vspace{0em}
\end{figure}

\subsection{Live inspection mode}

\begin{figure*}[t]
   \centering
   \includegraphics[width=\linewidth]{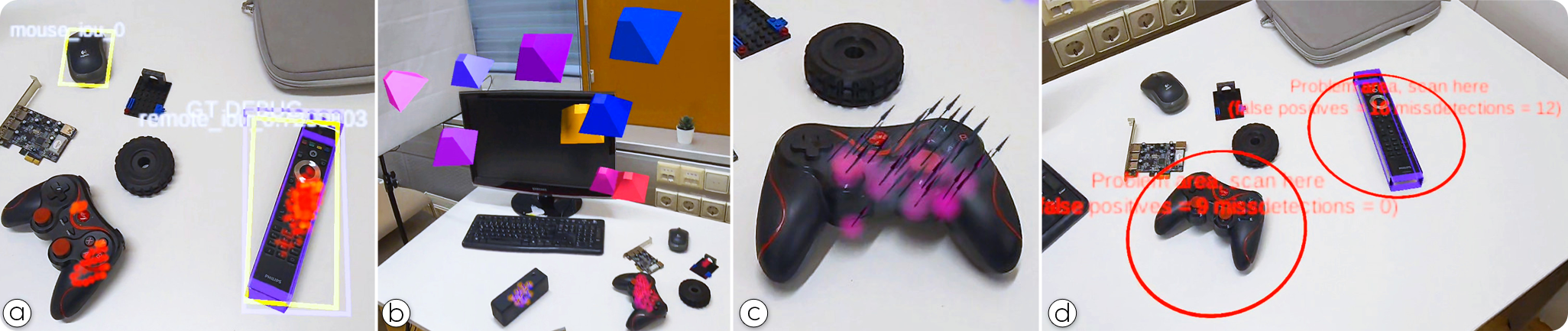}
  \caption{Visualizations of the situated explainability framework: (a) Registered explainability heatmaps from LayerCAM during live inference with detected bounding boxes (yellow) and annotated object (purple).(b) After the live inference, the 3D activation maps can be visualized per camera (c) In the zoomed view, the points have attached glyphs showing the directions of the cameras that captures them. (d) AR hints are shown where false positives or misclassifications happened.}
   \label{fig:visualizations}
    \vspace{0em}
\end{figure*}

\paragraph{Live inference}

At runtime, an AR visualization is based on the output of the visual model. The live webcam of the headset streams its output to a Python server that runs live network inference on a desktop computer, which applies the explainability methods. The output heatmaps are sent over the network to the AR application on the headset. Each point is raycast into the scene using the depth map of the headset for hits, and the explanations are displayed in AR directly at the object location, in a manner that is comprehensive to the users, as shown in Figure \ref{fig:teaser}a.

We use YOLOv8 and YOLOv12~\cite{tian2026yolov12} for object detection, and support RT-DETR~\cite{zhao2024detrs} and PaliGemma~\cite{beyer2024paligemma, steiner2024paligemma2} as alternatives for live mode. While YOLO is a one-stage object detector, RT-DETR is a transformer for object detection, and PaliGemma is a Vision-Language Model~(VLM), with natural language questions alongside images as input, and text as output. A screenshot of the VLM implementation is shown in Figure \ref{fig:example_vlm}.

\paragraph{Live visual explainability}

For the purpose of explainability, gradient activation methods are used in the last layers of the detection network. In our implementation, we use LayerCAM~\cite{jiang2021layercam}.
Compared to other CAM methods, the strength of LayerCAM is its localization in the image, being particularly suitable for applications where the user can interact with objects of different sizes. Visualizing the heatmaps from the CAM method answers the question ``for each image location, how important is this region for the model’s prediction of class $c$?'' Since the heatmaps are always displayed for one particular class, the target class $c$ needs to be specified.

The method is applied one frame at a time, with the highlighted pixels in the map signifying activations that are positively relevant to the target class. The latency introduced by the computation of the heatmap (four frames per second) suffices for real-time AR, since the inference is performed asynchronously and does not stall the AR experience (full headset frame rate).

At runtime, the direct output of the visual explainability is displayed in the 3D scene as red points that represent the intensity of the 2D maps (Figure \ref{fig:visualizations}a). To reduce clutter, the points fade out after a few seconds, but are saved for offline viewing. The colorization of the heatmap represents the relative importance of different regions for the model’s prediction of target class $c$. To handle a large number of points per frame, a grid filter is applied on each point cloud.

In addition to the explainability 3D heatmaps, backprojected bounding boxes of detected objects are displayed in the scene and can be toggled on or off. The areas outside the bounding boxes that are highlighted via heatmaps represent features relevant for the network, but not strong enough for full detections. For example, a controller (Figure \ref{fig:visualizations}a) has the features of a remote control (the buttons), but it is not detected as one. 

\paragraph{Ground truth}
\label{Sec:scene_setup}

While live inference does not necessarily need ground truth information, the framework offers the possibility to label objects by placing a 3D box around them and inputting their class label. This strategy has two benefits: First, the framework can offer information about wrong predictions by overlaying the real-time detected labels with ground truth, thereby identifying misclassifications, false positives or false negatives. Second, the framework can automatically record data annotated with ground truth. The annotation is performed by placing the boxes directly on top of the objects via hand interaction and scaling them along each axis (Figure \ref{fig:labelling}).

\subsection{Offline analysis mode}

The heatmaps from the live guidance are saved as point clouds together with the camera poses and detection information for all the processed frames of the session. In the following, we explain how to use this data for a series of explainability visualizations to gain more insights into a model's behavior. 

\paragraph{Spatially situated maps}
The spatially situated maps are the accumulated heatmap point clouds gathered in a live session, and are viewed offline, after the inference. All heatmap points are displayed with the same colorization as in live mode, directly on top of the workspace. Here, filtering is applied to combine points that overlap or are very close together. The users can also visualize all detections that occurred.

Alternatively, we implement a visualization that color-codes the points by the cameras from which they come, together with the camera displayed in its original pose (Figure \ref{fig:visualizations}b). This allows users to check which viewpoints cause a given activation and to identify problematic poses. Camera poses tend to be close together, and plotting all of them increases clutter. Hence, we cluster them by distance and orientation and group the point clouds accordingly. Cameras that are within a certain distance (15 centimeters) and rotation (15 degrees) are clustered together. Clusters of points can be individually deactivated to observe individual camera poses.

In practice, a particular viewpoint can lead to false positives or negatives. In such cases, the user might want to re-capture training data or consider what object features visible from a given point cause issues. Moreover, we display arrows that show the direction of the cameras from which the points were captured to help users visually identify the viewing directions (Figure \ref{fig:visualizations}c).

\paragraph{Performance analysis and visual AR hints}

To offer more high-level suggestions to the user, we integrate AR hints into the interface, highlighting areas in the scene that lead to problematic AI model behavior. This kind of visualizations is meant to give a broader impression of the session performance. 

For our implementation using object detectors with bounding box output, we overlap bounding boxes to compute correctness via Intersection over Union (IOU). The bounding box of the ground truth (Section \ref{Sec:scene_setup}) is projected onto the image, and the IOU is computed with each detection. By setting an IOU threshold and checking if the ground truth class is the same as the predicted class, detection errors can be identified as follows: 
True positives of target class \emph{c} are instances with matching bounding boxes that are in the same class. Misclassifications are instances with matching bounding boxes that are not in the same class. False positives are predictions of class \emph{c} not matching the ground truth, whereas a false negative means that correct detections of class \emph{c} have not occurred in the frame.

For an overview of these results, we show suggestions for outlier point clusters as areal highlights (Figure \ref{fig:visualizations}d). These outliers indicate areas with false positives and areas with false negatives. The user can choose to further capture data in the problematic areas (Section \ref{sec:datacapture}). We also compute statistics of the live session (number of frames, false positives, false negatives, true positives) and display them on a text banner to provide an immediate and quantitative evaluation of the detection performance.

\subsection{Data capture}

\label{sec:datacapture}

The application includes a data capture feature. After the user has explored the behavior of the AI model via situated explainability, they can immediately capture training data. For the bounding box labels, we project the 3D ground-truth labels (Section \ref{Sec:scene_setup}) onto the image plane of the passthrough camera, using the intrinsic parameters provided by the headset. All corners of the 3D labels in world space are projected. Then, their maximum and minimum values in X and Y coordinates are set as corners of 2D bounding boxes (Figure \ref{fig:labelling}). These boxes serve as image annotations for object detection.

However, the labels can be larger than the object in the 2D image. To make them tighter, we apply an automatic segmentation with Segment Anything~\cite{kirillov2023segment} in each box, recalculating the bounding boxes considering the bounds of the segmentation mask. Performed in batches without user intervention, we process the data to be directly compatible for YOLO-style label training.

\subsection{Implementation details}

For our evaluated implementation, we use a ~\emph{Meta Quest 3} headset, which offers a passthrough mode with access to the RGB camera, Unity3D as game engine, and a pre-trained YOLO model. 

Gradient-based explainability methods need the output of a network layer as input. To capture more high-level, visually expressive features, we use late layers of the networks for visualization. We set the detection confidence to a rather low threshold (0.3) to capture more information. Our implementation is inspired by previous work on image-based explainability \cite{jacobgilpytorchcam, yoloexplainability}. YOLO runs in real-time, which makes it more suitable for AR applications than foundation models. We only use the class loss of the network, since this is relevant for our purpose of class discrimination, enabling the user to investigate the detection of one class at a time.

\section{User Study 1: Exploratory Evaluation}
\label{sec:pilot}

We conducted an exploratory study to get first insights into how users engage with situated explainability; especially whether the participants find our approach useful and feedback on the features of our implementation. Our study received ethics approval from our university.

\subsection{Participants and study design}

For our initial qualitative exploration, we recruited participants with AI experience, to obtain informed feedback from users familiar with the concepts and techniques underlying the application. We recruited six users through word-of-mouth and email advertising (five male, one female), with an average age of 29 years (std=4.6). The average experience with ML is 3.4 on a scale of 0 (``no experience'') to 4 (``expert''), with one participant reporting ``basic familiarity'', and all others, ``advanced'' or ``expert'' level of familiarity. All participants had some AR experience, and all but one had experience with explainability tools. All participants received candy as compensation for their time.

For our study, we chose scenarios where participants need to identify weaknesses of the neural network-based object detectors, specifically multiple objects with some visual similarity, to pose a challenge for the detectors.

\textbf{Scenario S1: cluttered desk.} Participants were shown a cluttered desk (Figure \ref{fig:visualizations}) and were asked to inspect the behavior of an object detector trained on the COCO dataset to detect a television remote.

\textbf{Scenario S2: home pantry.} Participants were told that an AR assistant, running on their headset, detects groceries in their pantry (Figure \ref{fig:user_study_conditions}) to make recipes and shopping recommendations. The app is based on an object detector, which sometimes fails when detecting salt or beans. The task was to inspect and understand the behavior of that particular class.

Each user went through a familiarization task with the user interface (UI) and all features (live inference, offline inspection, and data capture) in S1. After familiarization, they were instructed to inspect the behavior of another detector using the application in S2. We encouraged users to think aloud. At the end of the study, we conducted a semi-structured interview, inquiring about the overall experience, system features, their understanding of the detector behavior, and participant demographics.

\subsection{Results}
We summarize the qualitative results of the user study, grouped into the main themes addressed by the participants.

\paragraph{Using situated explainability}
Users generally appreciated the spatial nature of the application and the ability to move around objects, for example participant (p1) stated that \textit{``in 2D you never have all the angles, and I wouldn't be able to look around the controller to see if it focuses on something else''}. When recording data, participants noted \textit{``For the labeling, it really helped that there was this point of view, if you're just recording data, it is easier because you can also get into the nooks and cracks''} (p2). Participants appreciated the object-centric character of the application, as it \textit{``definitely helped to be overlaid with the real object''} (p6). 

Another participant stated that they would use the application for first person debugging in a scenario where reliability is critical: \textit{``It would be super helpful to debug if you're like a surgeon and you want to check how these things work in real life before actually trying it on a live patient''} (p4).

The real-time character of the application, with quick feedback, was emphasized: \textit{``while looking around it could give you a feel of how good the model is without having to take pictures''} (p4). Users commented on enjoying the AR application: \textit{``it's kind of fun''} (p2) and \textit{``it can give me immediate feedback, it was really cool''} (p1).

\paragraph{Explainability features}
The participants addressed the usefulness of live heatmaps. For example, one user stated, inspecting the cluttered desk, \textit{``with the heat maps, I was immediately able to see what could be interpreted as remote''} and commented that the heatmap helps with \textit{``the potential object to cause false positives''} (p5). Another user stated \textit{``When you see the live activation maps showing something [...] you would be intrigued and look more''} (p6). This behavior is observed across participants, as another user got closer to the beans can with the live heatmap enabled, commenting: \textit{``A lot of the heat map is where the text is so maybe there's some information there''} (p1), being encouraged to explore more. Unexpected model failures are discovered in live exploration, with a participant referring to a poster on the wall in scene S2: \textit{``As a machine learning person I thought you placed those images there to be an adversarial attack [...] then I realized it was just a misfiring''} (p3). 

Compared to conventional pipelines, users saw the application potential: \textit{``It's hard to improve the detection, because you don't know what the system is looking at with these conventional systems [...] it's usually hidden inside it, so it definitely makes sense to visualize the activation somehow and then capture the data''} (p6). The live situated explainability experience was found to be easier to interpret \textit{``subjective instead of raw numbers''} (p5).

The 3D arrows and the camera poses from the offline inspection were considered useful by participants to convey spatial information about the captured viewpoints: \textit{``When I can walk around, the camera markers are way more practical''} (p5), \textit{“I could see the camera and the point clouds with the arrows, it was just kind of confirmation, and if I needed to capture more [...] of the area that I covered''} (p1). The data recording feature intuitively follows inspection of the model behavior: \textit{``[looking at the heatmaps] this makes sense. I would record a data set to improve the detection''} (p6).

\paragraph{Trust and reliability}

Participants reported mixed impressions about trust, but stated that they understand the model behavior better when using our situated explainability approach: \textit{``I would trust it more because now I understand why and when it fails''} (p2), which was echoed by another participant: \textit{``in line with what I expected [...] seeing it with all the objects in the scene I think I trust it a bit more''} (p4). The explainability features led a user to \textit{``have my thoughts confirmed [...] at least made me understand it.''} (p1).

After discovering failures, a participant stated: \textit{“The objects in the background should not have been an issue [...] therefore, it made me trust the model less''} (p3). Having more information beyond detection was appreciated: \textit{``to have an explainable way on the detections would definitely be more trustworthy because it's not that digital anymore. It's more like something in between.''} (p6).

Regarding the reliability of the detection, users also commented on using the application for testing the limits of the models: \textit{``I think it would be super useful if the conditions change a lot. If it's hard to record everything and you want to live test the behavior of the model [...] usually data sets are super clean and super curated, while, when you're in an actual operating room it might be a bit different, so if I'd be a surgeon, I would want to try beforehand before operating on a patient''} (p4).

\paragraph{Contrasting ML/AI work experience}
Participants also reflected on their experience when working with ML pipelines, stating that they usually
\textit{``look at the metrics and then maybe inspect some false positives and false negatives''} (p6) or \textit{``I look at the metrics. If it's a VLM, I will try some weird examples''} (p4).

They contrasted using our approach with conventionally inspecting outputs on the desktop: \textit{“This is much more fun [than static images], because it’s much more far away from what I do, if I'm in a hurry, I might want to do the other one''} (p3). Another user mentioned \textit{``on the screen you can run more statistical tests''} (p4). 

However, the interactivity was highly appreciated: \textit{``Right now I am looking at plots of loss functions, testing the network on images or video. But like this, it is not possible to get immediate feedback. For example, to notice like this part of the scene is causing trouble''} (p5). Participants appreciated the visualizations that help them understand the detectors: \textit{``I will trust more because I was able to look at better visualizations than plots of loss functions''} (p5).

Furthermore, a user with industrial experience stated: \textit{``I definitely prefer to use this system than just having the images on my computer. [...] As an engineer I'm using the system and testing it so I would prefer this, because you can really get in the viewpoints''} (p6).

\paragraph{LLM integration for explanations}
Since we also considered integrating an LLM into the system for additional explanations similar to related work\cite{xu2023xair, kim2025explainable}, we asked users whether they would like textual explanations to accompany the visualizations.

Most of the participants responded negatively, mainly due to concerns about visual clutter. One participant stated, \textit{“AR is a visual medium and the more text you put into it the more it gets cluttered“} (p2), while another remarked, \textit{“I would not use text because there's potentially already clutter''} (p3). 

Participants also expressed concerns about the reliability of LLM-generated explanations and their tendency to hallucinate. As p3 said, \textit{“[an LLM] would get useless very quickly, because it will try to find issues where there are not“}. P4 noted, \textit{“LLMs tend to fabricate meaning of meaningless stuff, I don't know if its behavior would be aligned with the explainability map [...]. I would not trust it if it just looks at an image“}. Participant p5 stated \textit{“no, I wouldn't say more text. With an LLM, I would trust it less“}. 

Only two participants saw benefits in textual explanations from LLM, but only in addition to existing features (\textit{“In addition that would be cool yes, but not instead“} (p1)) or for supporting less experienced users (\textit{“maybe if you don't have technical background, and you don't know what's going on, it would make sense to also have some text, but I didn't need it because it's rather straightforward with the image heat map“}, (p6)).
 
\paragraph{Suggested improvements}
Participants took note of the complexity of the UI and its multitude of features, stating \textit{``there were too many buttons''} (p3), \textit{``The UI is a bit confusing [...] I would like if it was a bit more hierarchical, so stuff opens and closes''} (p4), \textit{``The introduction was really necessary, otherwise I would have been really confused'' }(p2). Also, HMD ergonomics were a concern during prolonged use, one participant said \textit{“If this would be integrated into glasses that are lightweight''}(p3), and another one stated, \textit{``I wouldn't use it really day-to-day given the headset weight''} (p4).

\subsection{Discussion and iterative improvements}
\label{sec:iterative_improvements}

\paragraph{Situated explainability in AR} The application received positive feedback, with users appreciating the freedom to move in space, inspecting the environment, and receiving real-time feedback and 3D overlays. Both the real-time and offline modes showed different strengths, such as the immediate feedback, the directional arrow and camera visualizations, as well as the possibility of data recording. 

\paragraph{User interface improvements}
In response to user feedback, we simplified the UI layout to reduce clutter. In addition, we separated the functionality more clearly between live mode, offline mode, and data recording. 

\paragraph{User study condition} 
To compare with a conventional method for visual explainability, we decided to conduct another study. We added a second condition, namely visually inspecting pre-recorded static images from a test set (as mentioned by the users), and compared it to our situated explainability application. 

\paragraph{LLM addition} Overall, participants were skeptical about integrating LLM-generated textual explanations into the application. As all participants had ML expertise, the feedback suggests that expert users prioritize faithful and directly interpretable visual explanations over natural-language explanations that may introduce unsupported interpretations or hallucinations. While one participant acknowledged that textual explanations could benefit users with less technical background, another preferred it as an optional supplement. Therefore, we decided not to add an LLM as a feature to the tool.

\begin{figure}[t]
   \centering
   \includegraphics[width=\linewidth]{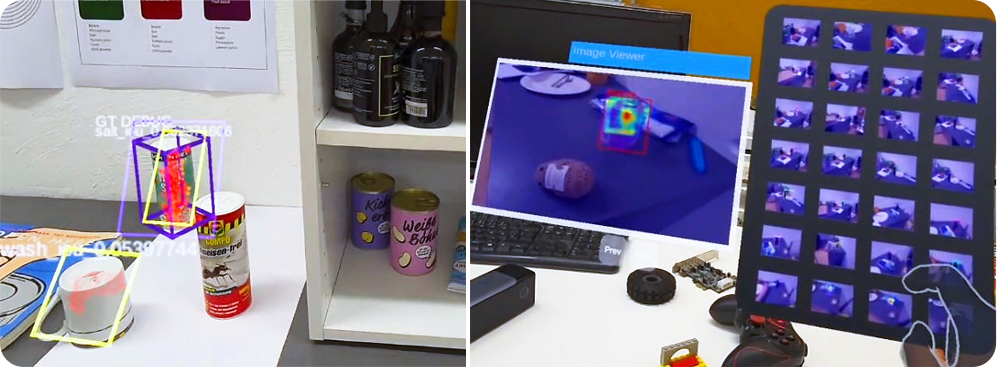}
  \caption{The 3D condition with situated explainability (left), and the 2D condition (right) as 2D user interface in AR, for inspection of explainability on a pre-recorded test set.}
    \label{fig:user_study_conditions}
    \vspace{0em}
\end{figure}

\section{User Study 2: Comparative User Study} 

Next, we aim to investigate how situated explainability is perceived in comparison to a well-established method: inspecting 2D images of a pre-recorded dataset with overlaid CAM maps.

\subsection{Participants and study design}

For the user study, we recruited people in the same way as for the first study in Section \ref{sec:pilot}, and we only targeted users who had experience with ML/AI. Ethics approval was obtained for this study from the ethics board of the local university. The study 
had 16 participants (none having participated in Study 1), with an average age of 31.44 (std=5.77), five females and 11 males. The familiarity with AI on a scale of 0 (``no experience'') to 4 (``expert'') was 3.13 on average, and the familiarity with explainability methods was 2.19, with all participants having at least ``basic familiarity''. Not all users were familiar with AR; two participants did not have experience at all, and the average was 1.69 on the scale from 0 to 4.

We ran a within-subject study, comparing the proposed application of situated explainability (study condition ``3D AR'') to inspecting network outputs with explainability heatmaps on a pre-recorded test set (study condition ``2D''). The users wore the AR headset in both conditions and could see the detection and the explainability method's output. Such heatmap explainability tools are currently used in practice on static, pre-recorded 2D images (Figure \ref{fig:user_study_conditions}).

Each study condition started with a familiarization phase, for users to get acquainted with the 2D and 3D interfaces. The familiarization phase of the 3D condition was again S1, while the familiarization phase of the 2D condition consisted of visually inspecting the output of a detector on a test set with cluttered objects on a table. For the two conditions, we used S2 with two different objects: a can of beans and a can of salt. The order of the conditions and of the two objects for the runs was randomized with a Latin square to mitigate learning effects. After each condition, the users completed NASA TLX, SUS \cite{brooke1996sus}, UEQ, and TiA \cite{korber2018theoretical} questionnaires. At the end of the study, we conducted a semi-structured interview about their experience, followed by a demographic questionnaire.

\subsection{Quantitative results}

For the analysis, normality was assessed using the Shapiro–Wilk test. When normality was satisfied, paired t-tests were used. Otherwise, Wilcoxon signed-rank tests were used. Holm correction was applied to families of related comparisons (NASA-TLX, UEQ, TiA subscales, and the correlations with the users' experience). Cohen's d is reported as the effect size.

To validate the within-subject design, we assessed whether presentation order influenced the results. Both the order of the interface and the order of the scenario were perfectly balanced between the participants (exact binomial tests, $p=1.000$). Furthermore, the preference for an interface was independent of the presentation order (Fisher's exact test, $p=1.000$). Although a nominal order effect was observed for the UEQ hedonic difference before correction ($p=.044$), no significant order effects remained after Holm correction. This indicates that the observed differences between the two conditions were not driven by presentation order or learning effects.

\paragraph{User experience, usability and mental load}

\begin{figure}[t]
    \centering
    \includegraphics[width=\linewidth]{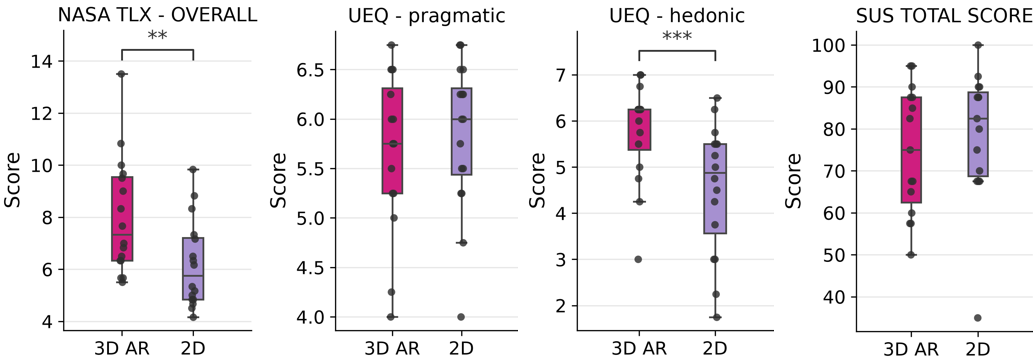}
    \caption{Result plots of the quantitative evaluation with the two conditions: 3D AR interface and 2D static images.}
    \label{fig:plots_1}
\end{figure}

The SUS scores for both conditions fall within the range of ``Good Usability'' (grade B), with average scores of 78.67 ($SD=15.86$) for the 2D condition, and 74.83 ($SD=15.04$) for the 3D condition (see Figure \ref{fig:plots_1}). The difference was not statistically significant (paired $t(15)=-1.05$, $p=.312$, Cohen's $d=-0.25$), indicating that the AR application achieved usability comparable to that of a well-established workflow.  
The UEQ did not reveal significant differences in pragmatic quality (paired $t(15)=-0.73$, $p=.476$, $d=-0.18$). In contrast, the hedonic quality of the 3D condition was rated significantly higher (paired $t(15)=4.18$, $p<.001$, $d=1.00$). Since the normality assumption was violated for the hedonic scores, the result was confirmed using a Wilcoxon signed-rank test ($p<.001$) and remained significant after Holm correction ($p=.002$). While the pragmatic scores indicated that participants perceived both systems as generally useful and effective in completing the task, the AR system was perceived as significantly more engaging. 

Regarding NASA-TLX, the 2D condition resulted in a significantly lower overall workload (2D: $M=6.19$, $SD=1.70$; 3D: $M=8.02$, $SD=2.25$; paired $t(15)=3.99$, $p=.001$, $d=0.92$). The result was confirmed by the Wilcoxon signed-rank test ($p=.002$). When examining individual dimensions, participants reported significantly higher mental demand ($p=.016$) and effort ($p=.007$) for the AR tool. However, after Holm correction for multiple comparisons, only the difference in effort remained statistically significant ($p=.044$). No significant differences were observed for physical demand, temporal demand, performance, or frustration. Therefore, users did not perceive that they performed better with the 2D interface, nor did they report differences in physical demand. 

\paragraph{User background experience impact}

To better understand the factors influencing user perception, we investigated whether participants' prior experience with AR, AI/ML, and explainability affected usability, user experience, workload, or trust. We hypothesized that familiarity with AI/ML and explainability could influence trust by enabling users to better interpret the presented explanations, while prior AR experience might reduce cognitive effort and correlate with measured usability associated with interacting in an immersive environment.

Participants with greater prior familiarity with AR did indeed rate the AR system as significantly more usable ($\rho=0.69$, adjusted $p=.014$), reported a significantly lower workload ($\rho=-0.66$, adjusted $p=.015$), and perceived the pragmatic quality of the AR interface more positively ($\rho=0.67$, adjusted $p=.024$). No significant relationships were observed between AR familiarity and the hedonic quality of the interface after Holm correction. Familiarity with AI/ML and explainability did not show any significant correlations with trust after Holm correction for multiple comparisons. 

\paragraph{Trust}

Overall, no significant differences were found between the two conditions. Participants rated the 2D interface slightly higher (2D: $M=3.72$, $SD=0.69$; 3D: $M=3.59$, $SD=0.80$), but the difference was not statistically significant (paired $t(15)=-0.70$, $p=.497$, $d=-0.15$). Similarly, no significant differences were observed for any of the other individual TiA subscales after Holm correction. Therefore, the presentation of explanations in AR neither increased nor decreased user trust compared to the 2D interface.

\paragraph{Preference}

Although the 3D interface was associated with a higher cognitive workload than the conventional desktop interface, participants showed a clear preference for the 3D interface. Thirteen of sixteen participants (81.3\%) preferred the 3D condition, while only three (18.8\%) preferred the 2D condition, a distribution that differed significantly from chance (exact binomial test, $p=.021$).

\subsection{Qualitative results}

For reporting qualitative results, we extracted direct user quotes from recorded sessions and grouped them into the most prevalent topics.

\paragraph{AR as medium for situated explainability}
Participants generally appreciated the \textbf{spatial nature of AR for explainability}, particularly the ability to explore the environment from different perspectives: \textit{“Because we live in a 3D world''} (p15), \textit{“You have more freedom to explore the environment [...] you know how the detection model works, and which angle is useful for the detection''} (p15). Compared to 2D visualizations, participants noted that \textit{“With 2D, much information is lost''} (p13) and appreciated \textit{“AR, because I can see not only the prerecorded data, but I can see from any perspective I want to see it“} (p12). This spatial interaction supported inspection: \textit{“very powerful that you can move around [...]. You can see very different zoom angles [...] Once you learn how it works, you can debug much quicker what are the problems of the model''} (p10). The spatially registered explanations themselves were considered easy to interpret: \textit{“The heat map mapped into the 3D space is very easy to interpret''} (p16).

A participant with experience in working with synthetic data particularly liked the AR condition because it is not simulated, allowing them to observe the \textbf{impact of the real environment}: \textit{``in AR you'll get an interaction with the real world, and you can anticipate where shadows fall realistically [...] It is the perfect mix, I've seen a lot of synthetically generated data sets, that's why''} (p11). Another participant stated \textit{``it gave me more ideas about how it works in real life''} (p20). The real-time character of \textbf{AR enables continuity over time}: \textit{``You get a sense for time consistency; that's a good feature about it, that's not something that you get from an image''} (p8).

Commenting on the innovative aspect of situated explainability, a user that was very experienced with AI explainability stated \textit{``It's really nice to see in real life the maps''} (p7), hinting at the practical value, with another user commenting on the novelty \textit{``nice way to debug a model, I haven't seen something like that before''} (p10).

\paragraph{Real-time interactivity and understanding}
Participants perceived the AR application to be \textbf{highly engaging}, one user stating \textit{``I see a huge potential for these interactive visualizations, for inspection, it really has a nice I-want-to-do-it touch to it [...] it's less boring''} (p11) and another user commenting \textit{``more intuitive and more engaging to see the stuff actually [...] if you see just images, you scroll through them, but you don't really engage with it, so I guess the engagement is really important in understanding stuff''} (p8). Users appreciated the ease of using the tool: \textit{``in a very natural way, you can scan the object you're interested in, otherwise you have to take a video with your mobile phone or the camera and process it, and you cannot move around''} (p16).

The registered 3D heatmaps help \textbf{to understand the AI behavior} in the pantry scenario (S2): \textit{``when you have this salt you can see that the object detection is more about the round object because every round object has some heatmap points on it, this model is maybe more about the size and the form of the objects''} (p12). A participant liked the \textit{``heat maps in general to look what is the model learning specially for the false positives, for example the dog food, it's looking at the title, because of the label, which is similar to the beans''} (p10). A user commented on how the overlaid heatmaps enhance understanding: \textit{``It's interesting to see what can confuse the model [...] like the buttons on the remote control, the buttons on the keyboard, I can easily understand this''} (p9).

Participants also noted that the 3D heatmaps could help \textbf{reveal model failures and draw attention to problematic cases}. One participant observed that the interactivity helped \textit{``see tipping points where the classifier suddenly fails, this might be way harder to detect with just images''}, while another user noted that \textit{``The heat map shows me where to look at, which additional images I should probably take care of''} (p11).

The freedom to move was found to be helpful: \textit{``if you change the camera position, you change the features of the 2D image''} (p9) providing a \textit{``really good intuition of what is happening''} (p8), while the 3D arrows showing the camera directions of the heatmaps are \textit{``interesting to see if there is a directional bias''} (p8). Furthermore, \textbf{fast decision-making} was enabled with live situated explainability \textit{``you don't have your fixed scenario and can adapt quickly''} (p18).

Users appreciated the recording feature: \textit{``nice that you can record the data very fast. I was sitting at the computer labeling data sets and it takes ages''} (p12), \textit{``in the real world, we can collect data sets everywhere''} (p15), or \textit{``from a video you cannot control it [...] [with 3D] I can collect the image I want''} (p15).

In contrast to standard tools, using situated explainability, a user stated that \textit{``The advantage is that you are the camera [..] You do two things at the same time, with one action you can classify an image and observe''} (p10), highlighting the real-time feedback. Users addressed \textbf{seeing the world from the AI model's perspective}: \textit{``The rendering of the features help most to see what the classifiers think (p8)''}, \textit{``see what the system sees''} (p22).

\paragraph{Limitations}

Users were generally content with the application features. However, two participants addressed headset-related drawbacks: \textit{``you cannot work three hours with this on your head''} (p16) or \textit{``I need to get used to see through the glasses''} (p9). In terms of other improvements, bounding box filtering, more numerical results, and visualizing prediction confidence were mentioned.

\paragraph{Trust calibration}

During the interviews, users again reported mixed experiences with trust, falling back to talking about 'understanding'. Some users stated that \textbf{deeper insights from the AR tool} make them understand more and therefore increase trust, because they \textbf{know how the network behaves}: \textit{``I understood what it did when I was running it, so I would say I would trust more after using it''} (p14). Other users stated: \textit{``Trust more in the sense that I have a better feeling or idea about its limitations''} (p11) or \textit{``In 3D, you can change the viewpoint, so I get more trust''} (p7). Other users addressed transparency increasing trust: \textit{``I trust it more because you can see what the system is seeing with the heatmaps''} (p12).

On the other hand, another user stated that \textbf{a deeper understanding and insight about failure} cases causes less trust, since they realized that the network can sometimes fail: \textit{``made me trust less because I discovered more problems''} (p10). Some users do not trust AI in general: \textit{``I do not trust AI [...] it helped to understand why it performed in this way or another way, but not to trust''} (p9).

\paragraph{Use-cases and applications}

As applied use-cases, users most often mentioned industry, manufacturing and medicine, for example: \textit{``in an industry setting this would be very beneficial''} (p16), using the AR application to check that a \textit{``component of the car is detected correctly from all important angles''} (p16). Other participants exemplified \textit{``screws, gears, mechanical direction, maybe, to find correct electric parts like resistors or capacitors''} (p13) or \textit{``identifying cracks on rails''} (p10). Possible applications were seen in \textit{``garment classification, that would be really helpful [...] any small scale industrial applications like food sorting where you can really walk around the scene [...] where a classification or segmentation model fails or comes up with a decision''} (p11). Another participant brought up \textit{``defect detection, I'm working now with this, so maybe if the system can find the feature of the good item, it could also find defect items''} (p9). The medical domain was addressed: \textit{``in a medical setting [...] for body parts''} (p16) or \textit{``medical cases, if you need to verify what is suggested by the network [on a patient]. If you have this tool to analyze the behavior, it might be a benefit''} (p7).
Broader use cases were also mentioned: \textit{``anything that you can inspect and objects you can touch''} (p10), highlighting the physicality of situated explainability, and also \textit{``action recognition: cooking, detect if you are cutting, stirring.''} (p22), hinting at dynamic scenarios.

\subsection{ Discussion of results}

\paragraph{Mental workload with situated explainability}
The workload advantage of the 2D interface appears to stem primarily from reduced cognitive burden rather than improved task success. This result is likely due to the fact that AR requires users to understand a three-dimensional representation, interact with a larger set of features, navigate more information simultaneously, and learn potentially unfamiliar interaction techniques. However, users overcame these challenges, subjectively reporting increased engagement, interest, and enjoyment in using situated explainability.

\paragraph{The impact of AR work experience}
Investigating the background of the study participants, the correlations of the AR experience with higher usability, lower workload, and higher pragmatic quality suggest that prior experience with AR can substantially influence how users perceive and interact with immersive explainability interfaces. This indicates rather the opposite of a novelty effect: one possible interpretation is that having experience with AR lets users focus on the actual functionality of the tool, making it easier for them to use it. Users were less distracted by the need to adjust and learn how to interact and operate in AR.

\paragraph{Preference for situated vs. conventional explainability}
Only three users preferred having the conventional 2D interface for inspecting a model's behavior via visual explainability. The most commonly cited reason for preferring the 2D interface was familiarity: \textit{``because I'm more used to that''} (p14) or \textit{``for speed and laziness I would prefer 2D, but overall having this live view and seeing the classes dependent on view angle and how you view it is much more valuable information''} (p13). Another user stated, \textit{``I'd prefer the live one if you could do an online fine-tuning''} (p17), suggesting network training during the application usage as an improvement.

\section{Discussion and future work}

\subsection{Lessons learned from the user evaluations}

\paragraph{The nature of debugging via situated explainability} With situated explainability, users move, get close, engage, and inspect the objects, enjoying the real-time multi-view capabilities and stating that they \textbf{get a better understanding of the AI models this way}. Interactivity, spatial movement, and direct coupling with the 3D world are essential to enable this experience. This insight is supported not only by subjective feedback but also by the majority of users who prefer the AR tool for AI model inspection. 

\paragraph{Engagement vs. mental load} Participants are willing to spend more mental effort because the AR experience is \textbf{perceived as more engaging and rewarding}. As indicated by the quantitative results and extensive user feedback, users received the situated explainability tool very positively. The desktop interface optimizes efficiency, while the AR interface optimizes engagement and exploration.

\paragraph{Trust vs. understanding} Throughout the evaluation, users referred to understanding when talking about trust, reporting mixed impressions about the increase in trust, explaining why the results of the TiA questionnaire were not significant. Mixed user feedback on trust is consistent with findings that show that systems with high transparency and low reliability decrease trust \cite{kaltenbach12017dual}. Our observations are similar to the results of de Heuvel et al. \cite{de2025immersive} about spatially grounded visual explanations in VR that improved users' objective understanding and perceived predictability of robot decisions, although no significant improvement in trust was observed. We believe that this topic warrants a larger discussion about working with flawed systems and trust, 
as literature shows that the accuracy of the system might be directly correlated with trust \cite{papenmeier2022s}.

\paragraph{Envisioned applications} The scenarios and applications addressed by the participants reflect what we envisioned in Section \ref{sec:situated_explainability}, namely that situated explainability enables understanding the behavior of the AI model in the real world, in tasks such as fault inspection or data bias discovery, and supports fast situated decision making, as well as informed data capture. Potential application domains span industrial, medical, and home applications, but also more specialized use-cases such as the understanding of adversarial attacks.

\subsection{Limitations and future work}

While revealing valuable user feedback, in our user studies, situated explainability was only tested by a limited number of users with AI/ML background. It would be interesting to explore how much understanding novices gain by interacting with the proposed tool and what level of help with interpretation would be needed. When recruiting participants with professional experience in ML/AI, it proved particularly challenging to achieve adequate gender balance. This situation is consistent with the gender imbalance among AI engineers: Women comprised approximately 27\% of individuals with AI engineering skills worldwide in 2023 \cite{oecd2024aiengineeringgender}.

In the proposed framework, only static objects are supported for the offline mode, since the heatmaps are accumulated, and objects are not tracked. In a future extension, we will aim for fully movable objects, which implies model-free 4DOF tracking.

Generally, situated explainability is not restricted to object detection and not even to image input. The pipeline could be easily adapted and integrated for object classification, or other tasks, by sending a different heatmap to the AR application and adapting the interface to the desired scenario. 

With the proposed application, we present a proof-of-concept implementation of situated explainability. In the future, it is important to explore and to evaluate different interface representations and their impact on user experience.

\section{Conclusion}

In this paper, we introduce \emph{situated explainability} and present example applications that showcase how AR can be used as a tool for visual explainability, registered in a user's surroundings, enabling rapid and intuitive understanding of AI model behavior. Our user studies indicate that explainability in AR created a more engaging experience at the cost of higher cognitive workload compared to traditional explainability workflows, suggesting a trade-off between having a rich experience and operational efficiency. 

In the current context of the rapid and broad adoption of AI across many fields, our aim is to set a basis for \emph{situated explainability}, showing that AR can be used as a tool to enable us to better understand AI. We hope this method provides a starting point for using AR as a medium enabling transparency, better understanding, and debugging of AI solutions, making use of the fully user-centric and spatial character of the immersive technology. Situated explainability allows natural and real-time movement and interaction, for testing and understanding in real environments---an endeavor to use the very nature of AR to \emph{see through the eyes of the AI}.


\bibliographystyle{abbrv-doi-hyperref}

\bibliography{template}

@inproceedings{selvaraju2017grad,
  title={Grad-cam: Visual explanations from deep networks via gradient-based localization},
  author={Selvaraju, Ramprasaath R and Cogswell, Michael and Das, Abhishek and Vedantam, Ramakrishna and Parikh, Devi and Batra, Dhruv},
  booktitle={Proceedings of the IEEE International Conference on Computer Vision},
  pages={618--626},
  year={2017}
}

@article{lundberg2017unified,
  title={A unified approach to interpreting model predictions},
  author={Lundberg, Scott M and Lee, Su-In},
  journal={Advances in Neural Information Processing Systems},
  volume={30},
  year={2017}
}

@inproceedings{muhammad2020eigen,
  title={Eigen-cam: Class activation map using principal components},
  author={Muhammad, Mohammed Bany and Yeasin, Mohammed},
  booktitle={2020 International Joint Conference on Neural Networks (IJCNN)},
  pages={1--7},
  year={2020},
  organization={IEEE}
}

@article{jiang2021layercam,
  title={Layercam: Exploring hierarchical class activation maps for localization},
  author={Jiang, Peng-Tao and Zhang, Chang-Bin and Hou, Qibin and Cheng, Ming-Ming and Wei, Yunchao},
  journal={IEEE Transactions on Image Processing},
  volume={30},
  pages={5875--5888},
  year={2021},
  publisher={IEEE}
}

@inproceedings{xu2023xair,
  title={Xair: A framework of explainable ai in augmented reality},
  author={Xu, Xuhai and Yu, Anna and Jonker, Tanya R and Todi, Kashyap and Lu, Feiyu and Qian, Xun and Evangelista Belo, Jo{\~a}o Marcelo and Wang, Tianyi and Li, Michelle and Mun, Aran and others},
  booktitle={Proceedings of the 2023 CHI Conference on Human Factors in Computing Systems},
  pages={1--30},
  year={2023}
}

@article{kundu2025relaxvr,
  title={RelaxVR: Cybersickness Reduction in Immersive Virtual Reality Through Explainable AI and Large Language Models},
  author={Kundu, Ripan Kumar and Hoque, Khaza Anuarul},
  journal={IEEE Access},
  year={2025},
  publisher={IEEE}
}

@article{castelo2023argus,
  title={Argus: Visualization of ai-assisted task guidance in ar},
  author={Castelo, Sonia and Rulff, Joao and McGowan, Erin and Steers, Bea and Wu, Guande and Chen, Shaoyu and Roman, Iran and Lopez, Roque and Brewer, Ethan and Zhao, Chen and others},
  journal={IEEE Transactions on Visualization and Computer Graphics},
  volume={30},
  number={1},
  pages={1313--1323},
  year={2023},
  publisher={IEEE}
}

@inproceedings{srinidhi2024xair,
  title={Xair: An xr platform that integrates large language models with the physical world},
  author={Srinidhi, Sruti and Lu, Edward and Rowe, Anthony},
  booktitle={2024 IEEE International Symposium on Mixed and Augmented Reality (ISMAR)},
  pages={759--767},
  year={2024},
  organization={IEEE}
}

@inproceedings{stanescu2023state,
  title={State-aware configuration detection for augmented reality step-by-step tutorials},
  author={Stanescu, Ana and Mohr, Peter and Kozinski, Mateusz and Mori, Shohei and Schmalstieg, Dieter and Kalkofen, Denis},
  booktitle={2023 IEEE International Symposium on Mixed and Augmented Reality (ISMAR)},
  pages={157--166},
  year={2023},
  organization={IEEE}
}

@inproceedings{stanescu2024error,
  title={Error Management for Augmented Reality Assembly Instructions},
  author={Stanescu, Ana and Mohr, Peter and Thaler, Franz and Kozinski, Mateusz and Skreinig, Lucchas Ribeiro and Schmalstieg, Dieter and Kalkofen, Denis},
  booktitle={2024 IEEE International Symposium on Mixed and Augmented Reality (ISMAR)},
  pages={690--699},
  year={2024},
  organization={IEEE}
}

@inproceedings{schieber2024asdf,
  title={Asdf: Assembly state detection utilizing late fusion by integrating 6d pose estimation},
  author={Schieber, Hannah and Li, Shiyu and Corell, Niklas and Beckerle, Philipp and Kreimeier, Julian and Roth, Daniel},
  booktitle={2024 IEEE International Symposium on Mixed and Augmented Reality (ISMAR)},
  pages={190--199},
  year={2024},
  organization={IEEE}
}

@inproceedings{dogan2024augmented,
  title={Augmented object intelligence with xr-objects},
  author={Dogan, Mustafa Doga and Gonzalez, Eric J and Ahuja, Karan and Du, Ruofei and Cola{\c{c}}o, Andrea and Lee, Johnny and Gonzalez-Franco, Mar and Kim, David},
  booktitle={Proceedings of the 37th Annual ACM Symposium on User Interface Software and Technology},
  pages={1--15},
  year={2024}
}

@article{raikwar2024beyond,
  title={Beyond the wizard of oz: negative effects of imperfect machine learning to examine the impact of reliability of augmented reality cues on visual search performance},
  author={Raikwar, Aditya and Mifsud, Domenick and Wickens, Christopher D and Batmaz, Anil Ufuk and Warden, Amelia C and Kelley, Brendan and Clegg, Benjamin A and Ortega, Francisco R},
  journal={IEEE Transactions on Visualization and Computer Graphics},
  volume={30},
  number={5},
  pages={2662--2670},
  year={2024},
  publisher={IEEE}
}

@article{langlotz2014next,
  title={Next-generation augmented reality browsers: rich, seamless, and adaptive},
  author={Langlotz, Tobias and Nguyen, Thanh and Schmalstieg, Dieter and Grasset, Raphael},
  journal={Proceedings of the IEEE},
  volume={102},
  number={2},
  pages={155--169},
  year={2014},
  publisher={IEEE}
}

@inproceedings{kim2019learning,
  title={Learning not to learn: Training deep neural networks with biased data},
  author={Kim, Byungju and Kim, Hyunwoo and Kim, Kyungsu and Kim, Sungjin and Kim, Junmo},
  booktitle={Proceedings of the IEEE/CVF Conference on Computer Vision and Pattern Recognition},
  pages={9012--9020},
  year={2019}
}

@article{zhou2020fine,
  title={Fine-grained visual recognition in mobile augmented reality for technical support},
  author={Zhou, Bing and G{\"u}ven, Sinem},
  journal={IEEE Transactions on Visualization and Computer Graphics},
  volume={26},
  number={12},
  pages={3514--3523},
  year={2020},
  publisher={IEEE}
}

@inproceedings{su2019deep,
  title={Deep multi-state object pose estimation for augmented reality assembly},
  author={Su, Yongzhi and Rambach, Jason and Minaskan, Nareg and Lesur, Paul and Pagani, Alain and Stricker, Didier},
  booktitle={2019 IEEE International Symposium on Mixed and Augmented Reality Adjunct (ISMAR-Adjunct)},
  pages={222--227},
  year={2019},
  organization={IEEE}
}

@INPROCEEDINGS{wang2020,
  author={Wang, Jialin and Liang, Hai-Ning and Monteiro, Diego Vilela and Xu, Wenge and Chen, Hao and Chen, Qiwen},
  booktitle={2020 IEEE International Symposium on Mixed and Augmented Reality Adjunct (ISMAR-Adjunct)}, 
  title={Real-Time Detection of Simulator Sickness in Virtual Reality Games Based on Players' Psychophysiological Data during Gameplay}, 
  year={2020},
  volume={},
  number={},
  pages={247-248},
  doi={10.1109/ISMAR-Adjunct51615.2020.00071}}

@article{song2025enhancing,
  title={Enhancing Patient Acceptance of Robotic Ultrasound through Conversational Virtual Agent and Immersive Visualizations},
  author={Song, Tianyu and Pabst, Felix and Eck, Ulrich and Navab, Nassir},
  journal={IEEE Transactions on Visualization and Computer Graphics},
  year={2025},
  publisher={IEEE}
}

@misc{jacobgilpytorchcam,
  title={PyTorch library for CAM methods},
  author={Jacob Gildenblat and contributors},
  year={2021},
  publisher={GitHub},
  howpublished={\url{https://github.com/jacobgil/pytorch-grad-cam}},
}

@misc{liu2023llava,
      title={Visual Instruction Tuning}, 
      author={Liu, Haotian and Li, Chunyuan and Wu, Qingyang and Lee, Yong Jae},
      publisher={NeurIPS},
      year={2023},
}

@misc{gpt4v,
      title={OpenAI. Gpt-4v(ision) system card.}, 
      author={Liu, Haotian and Li, Chunyuan and Wu, Qingyang and Lee, Yong Jae},
      year={2023},
        howpublished={\url{https://cdn.openai.com/papers/GPTV_System_Card.
pdf}},
}

@article{team2023gemini,
  title={Gemini: a family of highly capable multimodal models},
  author={Team, Gemini and Anil, Rohan and Borgeaud, Sebastian and Alayrac, Jean-Baptiste and Yu, Jiahui and Soricut, Radu and Schalkwyk, Johan and Dai, Andrew M and Hauth, Anja and Millican, Katie and others},
  journal={arXiv preprint arXiv:2312.11805},
  year={2023}
}

@article{wang2024yolov10,
  title={Yolov10: Real-time end-to-end object detection},
  author={Wang, Ao and Chen, Hui and Liu, Lihao and Chen, Kai and Lin, Zijia and Han, Jungong and others},
  journal={Advances in Neural Information Processing Systems},
  volume={37},
  pages={107984--108011},
  year={2024}
}

@inproceedings{zhao2024detrs,
  title={Detrs beat yolos on real-time object detection},
  author={Zhao, Yian and Lv, Wenyu and Xu, Shangliang and Wei, Jinman and Wang, Guanzhong and Dang, Qingqing and Liu, Yi and Chen, Jie},
  booktitle={2024 IEEE/CVF Conference on Computer Vision and Pattern Recognition (CVPR)},
  pages={16965--16974},
  year={2024},
  organization={IEEE}
}

@article{grubert2016towards,
  title={Towards pervasive augmented reality: Context-awareness in augmented reality},
  author={Grubert, Jens and Langlotz, Tobias and Zollmann, Stefanie and Regenbrecht, Holger},
  journal={IEEE Transactions on Visualization and Computer Graphics},
  volume={23},
  number={6},
  pages={1706--1724},
  year={2016},
  publisher={IEEE}
}

@inproceedings{castellano2024using,
  title={Using LLMs to explain AI-generated art classification via Grad-CAM heatmaps},
  author={Castellano, Giovanna and Miccoli, Maria Grazia and Scaringi, Raffaele and Vessio, Gennaro and Zaza, Gianluca and others},
  booktitle={Proceedings of 5th Italian Workshop on Explainable Artificial Intelligence, co-located with the 23rd International Conference of the Italian Association for Artificial Intelligence, Bolzano, Italy},
  year={2024}
}

@article{tinauer2022interpretable,
  title={Interpretable brain disease classification and relevance-guided deep learning},
  author={Tinauer, Christian and Heber, Stefan and Pirpamer, Lukas and Damulina, Anna and Schmidt, Reinhold and Stollberger, Rudolf and Ropele, Stefan and Langkammer, Christian},
  journal={Scientific Reports},
  volume={12},
  number={1},
  pages={20254},
  year={2022},
  publisher={Nature Publishing Group UK London}
}

@inproceedings{kirillov2023segment,
  title={Segment anything},
  author={Kirillov, Alexander and Mintun, Eric and Ravi, Nikhila and Mao, Hanzi and Rolland, Chloe and Gustafson, Laura and Xiao, Tete and Whitehead, Spencer and Berg, Alexander C and Lo, Wan-Yen and others},
  booktitle={Proceedings of the IEEE/CVF International Conference on Computer Vision},
  pages={4015--4026},
  year={2023}
}

@inproceedings{kundu2025pilar,
  title={PILAR: Personalizing Augmented Reality Interactions with LLM-based Human-Centric and Trustworthy Explanations for Daily Use Cases},
  author={Kundu, Ripan Kumar and Ahmed, Istiak and Hoque, Khaza Anuarul},
  booktitle={2025 IEEE International Symposium on Mixed and Augmented Reality Adjunct (ISMAR-Adjunct)},
  pages={273--280},
  year={2025},
  organization={IEEE}
}

@inproceedings{mendez2023does,
  title={How Does Explainability Look in Hybrid User Interfaces?},
  author={M{\'e}ndez, Juli{\'a}n and Satkowski, Marc and Rzayev, Rufat},
  booktitle={2023 IEEE International Symposium on Mixed and Augmented Reality Adjunct (ISMAR-Adjunct)},
  pages={251--256},
  year={2023},
  organization={IEEE}
}

@article{bhati2025survey,
  title={A survey of post-hoc xai methods from a visualization perspective: Challenges and opportunities},
  author={Bhati, Deepshikha and Amiruzzaman, Md and Zhao, Ye and Guercio, Angela and Le, Tram},
  journal={IEEE Access},
  year={2025},
  publisher={IEEE}
}

@inproceedings{la2023state,
  title={State of the art of visual analytics for explainable deep learning},
  author={La Rosa, Biagio and Blasilli, Graziano and Bourqui, Romain and Auber, David and Santucci, Giuseppe and Capobianco, Roberto and Bertini, Enrico and Giot, Romain and Angelini, Marco},
  booktitle={Computer Graphics Forum},
  volume={42},
  number={1},
  pages={319--355},
  year={2023},
  organization={Wiley Online Library}
}

@article{haedecke2023scrutinai,
  title={ScrutinAI: A visual analytics tool supporting semantic assessments of object detection models},
  author={Haedecke, Elena and Mock, Michael and Akila, Maram},
  journal={Computers \& Graphics},
  volume={114},
  pages={265--275},
  year={2023},
  publisher={Elsevier}
}

@inproceedings{lehman2020archie,
  title={Archie: A user-focused framework for testing augmented reality applications in the wild},
  author={Lehman, Sarah M and Ling, Haibin and Tan, Chiu C},
  booktitle={2020 IEEE Conference on Virtual Reality and 3D User Interfaces (VR)},
  pages={903--912},
  year={2020},
  organization={IEEE}
}

@inproceedings{zhou2016learning,
  title={Learning deep features for discriminative localization},
  author={Zhou, Bolei and Khosla, Aditya and Lapedriza, Agata and Oliva, Aude and Torralba, Antonio},
  booktitle={Proceedings of the IEEE Conference on Computer Vision and Pattern Recognition},
  pages={2921--2929},
  year={2016}
}

@inproceedings{xu2024multimodal,
  title={Multimodal 3d fusion and in-situ learning for spatially aware ai},
  author={Xu, Chengyuan and Kumaran, Radha and Stier, Noah and Yu, Kangyou and H{\"o}llerer, Tobias},
  booktitle={2024 IEEE International Symposium on Mixed and Augmented Reality (ISMAR)},
  pages={485--494},
  year={2024},
  organization={IEEE}
}

@inproceedings{qian2022arnnotate,
  title={Arnnotate: An augmented reality interface for collecting custom dataset of 3d hand-object interaction pose estimation},
  author={Qian, Xun and He, Fengming and Hu, Xiyun and Wang, Tianyi and Ramani, Karthik},
  booktitle={Proceedings of the 35th Annual ACM Symposium on User Interface Software and Technology},
  pages={1--14},
  year={2022}
}

@inproceedings{taheritajar2025scalable,
  title={Scalable Object Detection in Mixed Reality Using Incremental Re-Training and One-Shot 3D Annotation},
  author={Taheritajar, Alireza and Benson, Jeffrey and Gibson, Anthony and Wilburn, Brandon and Zhao, Jieqiong and Orlosky, Jason},
  booktitle={2025 IEEE International Symposium on Mixed and Augmented Reality (ISMAR)},
  pages={582--592},
  year={2025},
  organization={IEEE}
}

@inproceedings{ganj2026ar,
  title={AR as an Evaluation Playground: Bridging Metric and Visual Perception of Computer Vision Models},
  author={Ganj, Ashkan and Zhao, Yiqin and Guo, Tian},
  booktitle={Proceedings of the ACM Multimedia Systems Conference 2026},
  pages={72--83},
  year={2026}
}

@article{kim2025explainable,
  title={Explainable XR: Understanding user behaviors of XR environments using LLM-assisted analytics framework},
  author={Kim, Yoonsang and Aamir, Zainab and Singh, Mithilesh and Boorboor, Saeed and Mueller, Klaus and Kaufman, Arie E},
  journal={IEEE Transactions on Visualization and Computer Graphics},
  year={2025},
  publisher={IEEE}
}

@inproceedings{korber2018theoretical,
  title={Theoretical considerations and development of a questionnaire to measure trust in automation},
  author={K{\"o}rber, Moritz},
  booktitle={Congress of the International Ergonomics Association},
  pages={13--30},
  year={2018},
  organization={Springer}
}

@article{brooke1996sus,
  title={SUS-A quick and dirty usability scale},
  author={Brooke, John and others},
  journal={Usability evaluation in industry},
  volume={189},
  number={194},
  pages={4--7},
  year={1996},
  publisher={London, England}
}

@inproceedings{kaltenbach12017dual,
  title={On the dual nature of transparency and reliability: Rethinking factors that shape trust in automation},
  author={Kaltenbach1, Elizabeth and Dolgov, Igor},
  booktitle={Proceedings of the Human Factors and Ergonomics Society Annual Meeting},
  volume={61},
  number={1},
  pages={308--312},
  year={2017},
  organization={SAGE Publications Sage CA: Los Angeles, CA}
}

@article{carofilis2023improvement,
  title={Improvement of accent classification models through grad-transfer from spectrograms and gradient-weighted class activation mapping},
  author={Carofilis, Andr{\'e}s and Alegre, Enrique and Fidalgo, Eduardo and Fern{\'a}ndez-Robles, Laura},
  journal={IEEE/ACM Transactions on Audio, Speech, and Language Processing},
  volume={31},
  pages={2859--2871},
  year={2023},
  publisher={IEEE}
}

@article{theissler2022explainable,
  title={Explainable AI for time series classification: a review, taxonomy and research directions},
  author={Theissler, Andreas and Spinnato, Francesco and Schlegel, Udo and Guidotti, Riccardo},
  journal={Ieee Access},
  volume={10},
  pages={100700--100724},
  year={2022},
  publisher={IEEE}
}

@article{van2024harnessing,
  title={Harnessing eXplainable artificial intelligence for feature selection in time series energy forecasting: A comparative analysis of Grad-CAM and SHAP},
  author={Van Zyl, Corne and Ye, Xianming and Naidoo, Raj},
  journal={Applied Energy},
  volume={353},
  pages={122079},
  year={2024},
  publisher={Elsevier}
}

@article{li2024gt,
  title={GT-CAM: Game theory based class activation map for GCN},
  author={Li, Yanshan and Shi, Ting and Chen, Zhiyuan and Zhang, Li and Xie, Weixin},
  journal={IEEE Transactions on Pattern Analysis and Machine Intelligence},
  volume={46},
  number={12},
  pages={8806--8819},
  year={2024},
  publisher={IEEE}
}

@article{sun2025explainable,
  title={Explainable artificial intelligence for medical applications: A review},
  author={Sun, Qiyang and Akman, Alican and Schuller, Bj{\"o}rn W},
  journal={ACM Transactions on Computing for Healthcare},
  volume={6},
  number={2},
  pages={1--31},
  year={2025},
  publisher={ACM New York, NY}
}

@article{tzionis2025review,
  title={A review of explainable AI methods and their application in manufacturing systems},
  author={Tzionis, Grigorios and Mouratidis, Prodromos and Kougka, Georgia and Gialampoukidis, Ilias and Vrochidis, Stefanos and Kompatsiaris, Ioannis and Vlachopoulou, Maro},
  journal={Discover Applied Sciences},
  volume={8},
  number={1},
  pages={52},
  year={2025},
  publisher={Springer}
}

@article{rong2023towards,
  title={Towards human-centered explainable ai: A survey of user studies for model explanations},
  author={Rong, Yao and Leemann, Tobias and Nguyen, Thai-Trang and Fiedler, Lisa and Qian, Peizhu and Unhelkar, Vaibhav and Seidel, Tina and Kasneci, Gjergji and Kasneci, Enkelejda},
  journal={IEEE Transactions on Pattern Analysis and Machine Intelligence},
  volume={46},
  number={4},
  pages={2104--2122},
  year={2023},
  publisher={IEEE}
}

@inproceedings{alqaraawi2020evaluating,
  title={Evaluating saliency map explanations for convolutional neural networks: a user study},
  author={Alqaraawi, Ahmed and Schuessler, Martin and Wei{\ss}, Philipp and Costanza, Enrico and Berthouze, Nadia},
  booktitle={Proceedings of the 25th International Conference on Intelligent User Interfaces},
  pages={275--285},
  year={2020}
}

@inproceedings{wang2021explanations,
  title={Are explanations helpful? a comparative study of the effects of explanations in ai-assisted decision-making},
  author={Wang, Xinru and Yin, Ming},
  booktitle={Proceedings of the 26th International Conference on Intelligent User Interfaces},
  pages={318--328},
  year={2021}
}

@article{spinner2019explainer,
  title={explAIner: A visual analytics framework for interactive and explainable machine learning},
  author={Spinner, Thilo and Schlegel, Udo and Sch{\"a}fer, Hanna and El-Assady, Mennatallah},
  journal={IEEE Transactions on Visualization and Computer Graphics},
  volume={26},
  number={1},
  pages={1064--1074},
  year={2019},
  publisher={IEEE}
}

@article{papenmeier2022s,
  title={It’s complicated: The relationship between user trust, model accuracy and explanations in AI},
  author={Papenmeier, Andrea and Kern, Dagmar and Englebienne, Gwenn and Seifert, Christin},
  journal={ACM Transactions on Computer-Human Interaction (TOCHI)},
  volume={29},
  number={4},
  pages={1--33},
  year={2022},
  publisher={ACM New York, NY}
}

@article{wang2023generalized,
  title={A generalized explanation framework for visualization of deep learning model predictions},
  author={Wang, Pei and Vasconcelos, Nuno},
  journal={IEEE Transactions on Pattern Analysis and Machine Intelligence},
  volume={45},
  number={8},
  pages={9265--9283},
  year={2023},
  publisher={IEEE}
}

@inproceedings{de2025immersive,
  title={Immersive Explainability: Visualizing Robot Navigation Decisions through XAI Semantic Scene Projections in Virtual Reality},
  author={de Heuvel, Jorge and M{\"u}ller, Sebastian and Wessels, Marlene and Akhtar, Aftab and Bauckhage, Christian and Bennewitz, Maren},
  booktitle={2025 34th IEEE International Conference on Robot and Human Interactive Communication (RO-MAN)},
  pages={1633--1639},
  year={2025},
  organization={IEEE}
}

@article{huang2022conceptexplainer,
  title={ConceptExplainer: Interactive explanation for deep neural networks from a concept perspective},
  author={Huang, Jinbin and Mishra, Aditi and Kwon, Bum Chul and Bryan, Chris},
  journal={IEEE Transactions on Visualization and Computer Graphics},
  volume={29},
  number={1},
  pages={831--841},
  year={2022},
  publisher={IEEE}
}

@article{tian2026yolov12,
  title={Yolov12: Attention-centric real-time object detectors},
  author={Tian, Yunjie and Ye, Qixiang and Doermann, David},
  journal={Advances in Neural Information Processing Systems},
  volume={38},
  pages={78433--78457},
  year={2026}
}

@online{oecd2024aiengineeringgender,
  author  = {{OECD}},
  title   = {2023 LinkedIn data on OECD.AI: Definitions for AI occupations are more specific, women in more AI jobs as career transitions to AI grow},
  year    = {2024},
  url     = {https://oecd.ai/en/wonk/2023-linkedin-data},
  urldate = {2026-08-09}
}

@article{beyer2024paligemma,
      title={{PaliGemma: A versatile 3B VLM for transfer}},
      author={Lucas Beyer and Andreas Steiner and André Susano Pinto and Alexander Kolesnikov and Xiao Wang and Daniel Salz and Maxim Neumann and Ibrahim Alabdulmohsin and Michael Tschannen and Emanuele Bugliarello and Thomas Unterthiner and Daniel Keysers and Skanda Koppula and Fangyu Liu and Adam Grycner and Alexey Gritsenko and Neil Houlsby and Manoj Kumar and Keran Rong and Julian Eisenschlos and Rishabh Kabra and Matthias Bauer and Matko Bošnjak and Xi Chen and Matthias Minderer and Paul Voigtlaender and Ioana Bica and Ivana Balazevic and Joan Puigcerver and Pinelopi Papalampidi and Olivier Henaff and Xi Xiong and Radu Soricut and Jeremiah Harmsen and Xiaohua Zhai},
      year={2024},
      journal={arXiv preprint arXiv:2407.07726}
}

@article{steiner2024paligemma2,
      title={{PaliGemma 2: A Family of Versatile VLMs for Transfer}},
      author={Andreas Steiner and André Susano Pinto and Michael Tschannen and Daniel Keysers and Xiao Wang and Yonatan Bitton and Alexey Gritsenko and Matthias Minderer and Anthony Sherbondy and Shangbang Long and Siyang Qin and Reeve Ingle and Emanuele Bugliarello and Sahar Kazemzadeh and Thomas Mesnard and Ibrahim Alabdulmohsin and Lucas Beyer and Xiaohua Zhai},
      year={2024},
      journal={arXiv preprint arXiv:2412.03555}
}

@inproceedings{white2009sitelens,
  title={SiteLens: Situated visualization techniques for urban site visits},
  author={White, Sean and Feiner, Steven},
  booktitle={Proceedings of the SIGCHI Conference on Human Factors in Computing Systems},
  pages={1117--1120},
  year={2009}
}

@misc{yoloexplainability,
  title={EigenCAM for YOLO 26 Interpretability},
  author={Rigved Shirvalkar},
  year={2021},
  publisher={GitHub},
  howpublished={\url{https://github.com/rigvedrs/YOLO-26-CAM}},
}


\end{document}